\documentclass[a4paper,fleqn,usenatbib]{mnras}

\usepackage{newtxtext,newtxmath}
\usepackage[T1]{fontenc}
\usepackage{ae,aecompl}

\usepackage{graphicx}	% Including figure files
\usepackage{amsmath}	% Advanced maths commands
\usepackage{amssymb}	% Extra maths symbols
\usepackage{subfig}
\usepackage{siunitx}

\title[CapsNets for morphological classification]{Galaxy morphology prediction using capsule networks}

\author[Katebi. et al]{
Reza Katebi$^{1}$\thanks{E-mail: \url{rk726014@ohio.edu}},
Yadi Zhou $^{2}$,
Ryan Chornock$^{1}$
and Razvan Bunescu$^{3}$
\\
$^{1}$Astrophysical Institute, Department of Physics and Astronomy, Ohio University, Athens, OH 45701, USA\\
$^{2}$Department of Chemistry and Biochemistry, Ohio University, Athens, OH 45701, USA\\
$^{3}$School of Electrical Engineering and Computer Science, Ohio University, Athens, OH 45701, USA 
}

\date{Accepted XXX. Received YYY; in original form ZZZ}

\pubyear{2015}

\begin{document}
\label{firstpage}
\pagerange{\pageref{firstpage}--\pageref{lastpage}}
\maketitle

% Abstract of the paper
\begin{abstract}
Understanding morphological types of galaxies is a key parameter for studying their formation and evolution. Neural networks that have been used previously for galaxy morphology classification have some disadvantages, such as not being invariant under rotation. In this work, we studied the performance of Capsule Network, a recently introduced neural network architecture that is rotationally invariant and spatially aware, on the task of galaxy morphology classification. We designed two evaluation scenarios based on the answers from the question tree in the Galaxy Zoo project. In the first scenario, we used Capsule Network for regression and predicted probabilities for all of the questions. In the second scenario, we chose the answer to the first morphology question that had the highest user agreement as the class of the object and trained a Capsule Network classifier, where we also reconstructed galaxy images. We achieved promising results in both of these scenarios. Automated approaches such as the one introduced here will greatly decrease the workload of astronomers and will play a critical role in the upcoming large sky surveys.

\end{abstract}

% Select between one and six entries from the list of approved keywords.
% Don't make up new ones.

%%%%%%%%%%%%%%%%%%%%%%%%%%%%%%%%%%%%%%%%%%%%%%%%%%

%%%%%%%%%%%%%%%%% BODY OF PAPER %%%%%%%%%%%%%%%%%%

\section{Introduction}
Morphological classifications have been used by astronomers to classify galaxies based on their visual aspects such as size, colour and shape. Studying morphological classifications is crucial to understand the evolution of galaxies and their properties such as age, formation and interaction with other galaxies. All-sky surveys are the key solutions to probe galaxy formation and evolution. 
 
 In order to conduct these studies, observation of a large number of galaxies and determination of their morphological classification is crucial. Large sky surveys such as the Sloan Digital Sky Survey (SDSS) \citep[e.g.,][]{blanton2017sloan} provided a large amount of data for the objects in our universe including galaxies. The morphological classification of galaxies has been traditionally done by experts, which is both inefficient and impractical for the large datasets available from current sky surveys and even larger upcoming ones such as the Large Synoptic Survey Telescope (LSST) \citep{ivezic2008lsst}. The Galaxy Zoo project \citep{lintott2008galaxy} started with the hope of partially solving this problem by a crowdsourcing method. The project was very successful and $\sim 900, 000$ galaxies were classified by online participants in a time span of months. Since then, other iterations of the Galaxy Zoo project  have annotated other datasets with more complex classification schemes \citep[e.g.,][]{willett2016galaxy}. However, even this approach is not feasible for the available and upcoming large datasets. 
 
 The amount of data is increasing as modern telescopes continue to take data, and projects like LSST will significantly increase the number of galaxies observed. Therefore, classifying these galaxies by crowdsourcing and visual inspection is next to impossible and developing an automated classification tool is necessary. Recently, improvement in computer vision techniques primarily through deep neural networks \citep[e.g.,][]{krizhevsky2012imagenet} and available computing power through GPUs have made this automated approach more promising. 
 
 In an attempt to find an automated classification approach, an international competition was launched by Galaxy Zoo on Kaggle \footnote{\url{https://www.kaggle.com/c/galaxy-zoo-the-galaxy-challenge}} using the images from the Galaxy Zoo 2 project \citep{willett2013galaxy} and the wining team provided a convolutional neural network (CNN) that exploits both translational and rotational symmetry in the images. This method can produce near perfect accuracy of $>99$\% for the images with a high agreement among the Galaxy Zoo participants \citep{dieleman2015rotation}. However, in order to extract the viewpoint, \cite{dieleman2015rotation} flipped, rotated and cropped images to extract 16 viewpoints for each image. Next, they trained a neural network with 16 convolutional neural subnetworks for each of the extracted viewpoints that are all connected to two fully connected layers. Each of these subnetworks learns the same features in a particular viewpoint that are useful for the classification task \citep{dieleman2015rotation}. However, the problem with this approach is that it cannot cover all of the possible rotations, orientations and their combinations; therefore, it still heavily depends on different training setups. Another problem is that this method is computationally very expensive because it relies on training multiple subnetworks. Moreover, it has been known that CNNs lose valuable information such as spatial hierarchies between features in the image. They also lack rotational invariance that causes CNNs to incorrectly assign labels to objects as long as a set of features is present during the test time disregarding the spatial relationship of these features to each other \citep{sabour2017dynamic}. 
 
 Recently, \cite{sabour2017dynamic} introduced a new type of network structure called Capsule Network (CapsNet) to address these issues in CNNs. This new structure contains capsules that are a nested set of layers. In contrast to traditional CNNs, this network is spatially aware and rotationally and transitionally invariant with the use of dynamic routing and reconstruction as regularization. \cite{sabour2017dynamic} achieved state of the art result of $0.25$\% test error on the Modified National Institute of Standards and Technology (MNIST) dataset of handwritten digits with shifting the images only by two pixels without applying any other data augmentation methods (e.g., rotation, flipping, scaling, etc.). In this work, we are proposing the use of CapsNet for the task of galaxy morphology prediction as a better alternative for CNNs. 
 
 \section{Galaxy Zoo 2}
 
The Galaxy Zoo is an online project where participants described galaxy morphology classification by answering a series of questions on the coloured images of the galaxies. In this work, we used data from the Galaxy Zoo 2 project where the participants answered 11 questions with 37 answers in total \citep{willett2013galaxy}. These questions were designed in a hierarchical manner where the next question was chosen based on the answer of the participant to the previous question. Each individual answered a subset of questions based on the way the decision tree was designed. The answers provided by users for one image transformed to a set of weighted vote fractions. These results have been used to study structure, formation and evolution of the galaxies \citep[e.g.,][]{skibba2009galaxy}. Also, the accuracy of the results from the Galaxy Zoo projects was confirmed by comparing them with smaller samples classified by experts and automated pipelines  \citep{bamford2009galaxy, willett2013galaxy}. Here, we used the dataset provided by Galaxy Zoo 2 for an international contest. The galaxies were selected with a variety of colours, sizes, and morphological classes. The goal of the project was to find an algorithm that could be applied to many different types of galaxies in the upcoming surveys. The total number of objects was limited by the depth of imaging in SDSS and the morphological categories that were over-represented as a function of the colour. This approach ensured that the colour does not play a role in the morphological classification and the models are purely based on the structures of the galaxies observed in the images. We only used the training set of the provided dataset because we did not have access to the labels of the validation dataset. The training set consisted of 61,578 JPEG coloured images of the galaxies with the size of $424 \times 424$ pixels. The morphological data was in the form of cumulative probabilities that gave higher weights to the questions that were asked higher in the question tree and determined a more fundamental morphological structure. 
 
 The goal of the contest was to predict probabilities for each of the 37 answers in the question tree; therefore, the task was a regression as opposed to classification. However, in this work we also reconstructed galaxy images based on the answers to question 1. This classification scheme is discussed in Section ~\ref{Reconst} in more detail. 

\section{Related Work}

Machine learning techniques such as neural networks have been used in astronomy research in the past few decades \citep[e.g., photometric redshift estimation;][]{collister2004annz, firth2003estimating}. Galaxy morphology classification is traditionally done by manually extracting a number of features that are known to discriminate different classes. Examples of these features are: surface brightness, ellipticity, concentration, radii, and log-likelihood values measured from different types of radial profiles \citep[e.g.,][]{storrie1992morphological}.

\cite{storrie1992morphological} used feed forward neural networks and 13 extracted parameters as input for training a classifier. Subsequent works used other machine learning methods such as kernel support vector machines (SVMs) \citep{tasca2009robust} and principal component analysis (PCA) \citep{naim1995automated, lahav1995galaxies, de2004machine} to extract features from the images. Next, they trained feed-forward neural networks using these features. These methods still heavily rely on feature extraction. In another approach, researchers used general purpose image features rather than galaxy-specific ones to perform galaxy morphology classification combined with nearest-neighbor classifiers \citep[e.g.,][]{kuminski2014combining}. 
 
Recently, \cite{dieleman2015rotation} used CNNs for this task. Their approach is different from the ones introduced before in two ways.  First, the morphological classification scheme provided by Galaxy Zoo 2 was a much more fine-grained task compared to the past work (mentioned above) where the task was classifying galaxies into a limited number of morphological classes \citep[except][]{kuminski2014combining}. Second, they did not use any prior handcrafted features or features that were extracted using machine learning algorithms such as PCA and SVM, which typically need many hours to develop. Instead, their proposed deep neural network learns hierarchies of features that allow the network to detect more abstract and complex features in the images. 

One important aspect of their approach is that their method exploits rotational and translational symmetry in the images. To do that, they constructed 16 different viewpoints for each image by rotating, cropping, and flipping the image. Next, they have used one CNN with 4 convolutional plus pooling layers for each of these viewpoints and connected all 16 CNNs to two fully connected layers that were regularized using dropout method \citep{hinton2012improving}. However, their method cannot cover all of the possible rotations, orientations, and their combinations; therefore, it still heavily depends on pre-training data manipulations. Moreover, there are disadvantages for using CNNs; specifically, they are known to lose important information about the spatial hierarchies between features in the image during the pooling process (usually max pooling) or, in other words, they are not spatially aware \citep{sabour2017dynamic}. In our approach, we used CapsNet, which was proposed to solve the problems of CNNs that were discussed above. CapsNet is rotationally and transitionally invariant because it uses a unique type of algorithm called ``routing by agreement'' and applies reconstruction as regularization. We will discuss the structure of CapsNet in the Section \ref{CapsNet} in more details. 

\section{Approach} 
 
In this section, we discuss our approach for galaxy morphology classification that is quite different from other ones proposed earlier. First, we discuss our experimental setup. Next, we talk about the preprocessing that we did in order to prepare the data for training. Last, we discuss the network structure, training process and our implementation.

\subsection{Experimental Setup} \label{ES}

The dataset that we used contains 65,578 images with associated portabilities of 37 answers of the questions asked during Galaxy Zoo 2 project. The task on the competition was to predict the probabilities for each of these 37 answers and calculate a root-mean-square-error (RMSE). We took two approaches here. In the first approach, we calculated RMSE, which was the goal of competition. In the second scenario, we took only the answers to the first question as the ground truth and chose objects where annotators had more than 0.8 agreement on choosing one answer where the participants chose among the first two answers. Therefore, we assigned two classes to the training examples based on the answer with the highest probability. In both evaluation scenarios, we divided the dataset to 80\% training and 20\% testing subsets.  

\subsection{Data Preprocessing}

We first cropped the images to reduce the dimensions of the input to the network. The majority of the objects were in the center of the images that fit in a square smaller than the size of the image; therefore, we cropped images from $424 \times 424$ pixels to $216 \times 216$ pixels and then down-sampled them 3 times to $72 \times 72$ pixels. We shifted images 2 pixels in each dimension with zero padding. We did not do any other data preprocessing and augmentation because CapsNet performs well with small datasets \citep{sabour2017dynamic}. We did not convert coloured images to grey scale because we observed that the accuracy is higher when using the coloured version and the reason behind it is that there is colour difference between the different parts of the galaxy such as bulge and the disk components. 

\subsection{Capsule Network}\label{CapsNet}
Capsules in CapsNet \citep{sabour2017dynamic} are groups of neurons that output vectors that represent different poses of the input. One of the disadvantages of the CNNs as mentioned before comes from pooling layers. In order to overcome this, CapsNet replaces pooling layers with an algorithm called ``routing by agreement''. In this algorithm, the lower layer capsules or Primary capsules predict the output of the next layer capsules or parent capsules. The routing weights get stronger if these predictions have a strong agreement with the actual outputs of the parent capsules and weaker if they disagree during the routing iterations. Taking $\^u_{i}$ as the activation function for the capsule $i$ in the layer $l+1$, the predicted output $\hat{u}_{j|i}$ of the capsules in the layer $l+1$ is represented by,
\begin{equation}
\hat{u}_{j|i} = W_{ij} u_{i}
\end{equation} 
where $W_{ij}$ is learned by the network during the backward propagation. Next, the coupling coefficients of the primary and parent capsules ($c_{ij}$) are calculated by applying a \textit{Softmax} function on the initial logits $b_{ij}$ that are set to zero at the initial stage of the routing by agreement process, 
\begin{equation}
c_{ij} = \frac{exp(b_{ij})}{\sum_{k} exp(b_{ik})}
\end{equation}  
where $k$ is the number of capsules in the next layer. After that, the input layer of the parent capsules $j$ in layer $l+1$ is calculated as follows:
\begin{equation}
s_{j} = \sum_{i} c_{ij} \hat{u}_{j|i}
\end{equation}
Then, a non-linear squashing function represented in eq.~\ref{squash} is applied on the output vectors to keep their length between 0 and 1 because the length of these vectors represent the probability of the presence of the object in the image,
\begin{equation}\label{squash}
v_{j} = \frac{|| s_{j}||^{2}}{1 + || s_{j}||^{2}} \frac{s_{j}}{||s_{j}||} 
\end{equation}
Next, the log probabilities are updated by the actual outputs of the $v_{j}$ capsules $j$ in layer $l+1$ and the predicted outputs $\hat{u}_{j|i}$ as following,
\begin{equation}
b_{ij} \leftarrow b_{ij} + v_{j} \cdot \hat{u}_{j|i}
\end{equation}
Each of the capsules $k$ in the last layer is associated with a loss function $l_{k}$ that has the following from, 
\begin{equation}
l_{k} = T_{k}max\Big(0, m^{+} - ||v_{k}||\Big)^{2} + \lambda(1-T_{k})max\Big(0, ||v_{k}|| - m^{-}\Big)^{2}
\end{equation}
where $T_{k}$ is one when class $k$ is present and zero otherwise. In this work, we chose $\lambda=0.5$, $m^{+}=0.9$ and $m^{-}=0.1$ for consistency with previous work \citep{sabour2017dynamic}.

In the case of regression, we used mean-square error (MSE) between the predictions and true crowd-sourced probabilities as the loss function that is as following,
\begin{equation}
MSE(p_{k}^{\prime} , p_{k})  = \sum_{k=1}^{37} (p_{k}^{\prime}  - p_{k})^{2}
\end{equation}
where $p_{k}$ is the answer probabilities associated with an image and $p_{k}^{\prime}$ are probabilities predicted by the network.

\subsection{Network Architecture}\label{arc}
\subsubsection{Baseline Network}
For the classification schemes, we used a standard CNN model with the following structure:
\begin{itemize}
	 \item $72 \times 72$ downsampled images of the galaxies as input.
	 \item A convolution layer with 512 filters with a receptive field of $9\times9$ and a stride of 1.
 	\item Max pooling with a receptive field of $2\times2$ and a stride of 2.
 	\item Rectified Linear Function (ReLU) \citep{glorot2011deep, nair2010rectified} as activation function
 	\item A convolution layer with 256 filters with a receptive field of $5\times5$ and a stride of 1.
 	\item Max pooling with a receptive field of $2\times2$ and a stride of 2.
 	\item  ReLU as activation function
 	\item A fully connected layer with 1024 neurons with ReLU as their activation function followed by the dropout rate of 0.5.
 	\item A fully connected layer with 1024 neurons with ReLU as their activation function followed by the dropout rate of 0.5.
 	\item A fully connected layer with Log-Softmax as their activation function where the number of neurons is assigned based on the number of classes in the classification scheme.
\end{itemize}

For the baseline network we used negative log-likelihood as the loss function for the classification scheme and we removed the the last fully connected layer, dropout and the Log-Softmax layer when calculating RMSE in the regression scheme. 

\subsubsection{Capsule Network}
\begin{figure*}
	\includegraphics[scale=0.5]{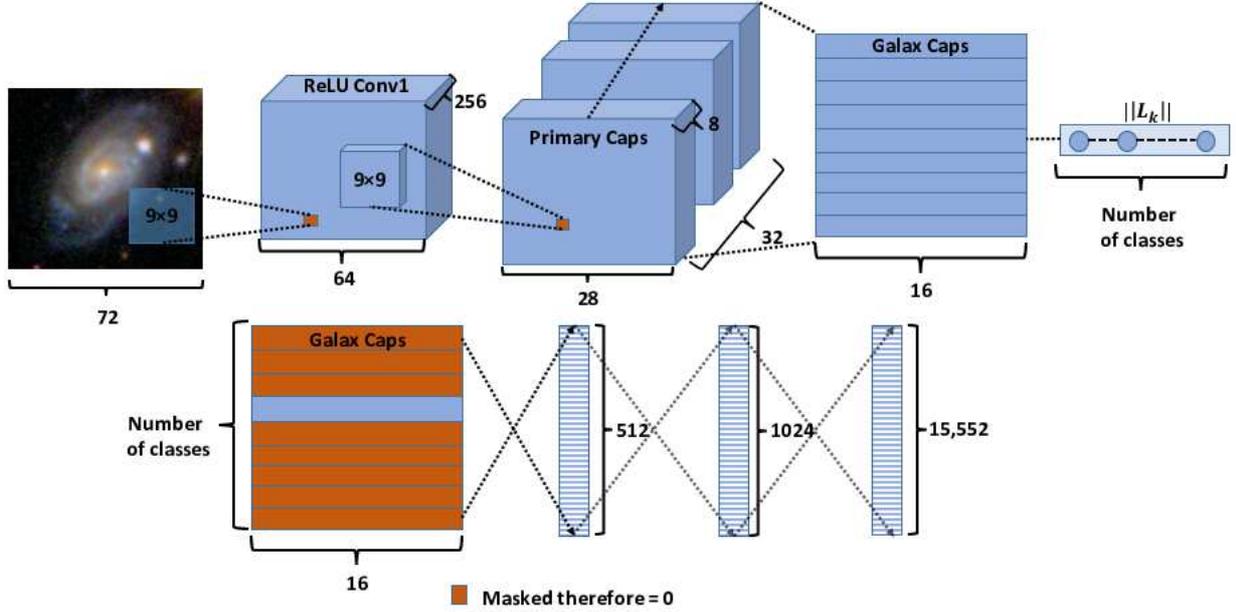}
    \caption{The architecture of the model used in this work. \textit{Top:} represents architecture of the capsule layers where the GalaxCap layer has 2 or 37 capsules based on the different setups discussed in Section \ref{arc}. \textit{Bottom:} represents the structure of the decoder that acts as regularization during the training}
    \label{Capsules}
\end{figure*}

Our network structure was based on the original CapsNet introduced by \cite{sabour2017dynamic} with some minor changes because of the size of the input images that is shown in Figure~\ref{Capsules}. The structure of the network was as following:
\begin{itemize}
	\item{Inputs:} $72 \times 72$ downsampled images of the galaxies.
	\item{Layer 1:} a convolutional layer with 256 filters with a receptive field of $9\times9$ and a stride of 1 with no zero padding that lead to the 256 feature maps with the size of $64 \times 64$.
	\item{Layer 2:} second convolutional layer with 256 filters with a receptive field of $9\times9$ and a stride of 2 applied and then reshaped to 32 primary capsules with 8 dimensions where each dimension is a feature map with the size of $28 \times 28$.
	\item{Last layer:} 2 or 37 capsules based on the training scheme studied in this work where each of them represented one class.
	\item{Decoder:} the decoder part of the network was composed of three fully connected layers with 512, 1024 and 15,552 neurons respectively where the neurons in the first two had ReLU as their activation function and the neurons of the last layer had a Sigmoid activation function. The number of neurons in the last layer were equal to the number of pixels in the input image. In fact, the reconstruction loss is the squared difference between the reconstructed image and the input image and it was scaled to 0.0005, so it would not dominate during the training process. 
\end{itemize}

The decoder part of the network forces the capsules to learn features during the training that are useful for the reconstruction of the image; therefore, it acts like a regularization for the network and controls the overfitting. For the regression task, we removed the decoder part of network and computed RMSE as discussed in Section~\ref{CapsNet}.

\subsection{Implementation and Resources}  
We implemented our model \footnote{\url{https://github.com/RezaKatebi/Galaxy-Morphology-CapsNet}} in Python using the Pytorch library based on the code provided in \cite{gram-ai_2018} that enabled us to use GPU acceleration. Moreover, the Pytorch library  carried out the differentiations with the \textit{autograd} method. We used one NVIDIA Tesla P100 GPU unit along with 4 CPUs on the Owen cluster at the Ohio Supercomputer Center (OSC) with 16Gb of memory \citep{OhioSupercomputerCenter1987}. For training our networks, we used an Adam optimizer. 

\section{Results}

\subsection{Regression}
\begin{figure*}
	\centering
 	\subfloat[Training]{\includegraphics[width=0.49\textwidth]{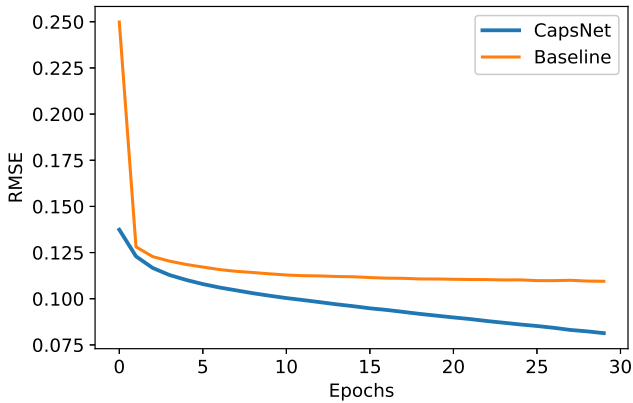}}\hfil 
 	\subfloat[Testing]{\includegraphics[width=0.49\textwidth]{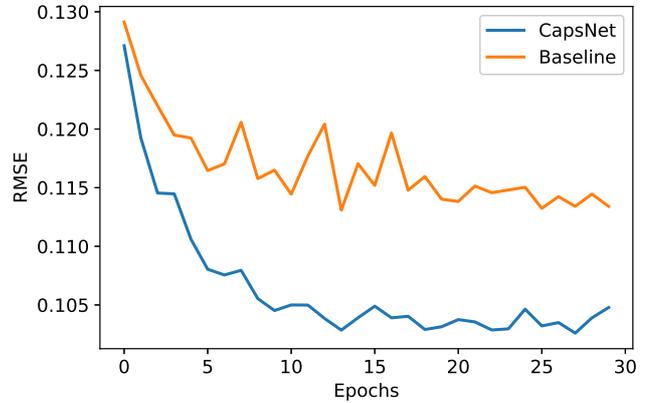}}
    \caption{Training and testing RMSE vs number of epochs for classification based on the answers to question one.}
    \label{fig_rmse}
\end{figure*}

In this section, we removed the decoder part of the CapsNet and computed the RMSE between the predictions and true crowd-sourced probabilities as explained in  Section ~\ref{CapsNet}. We also removed the last fully connected layer, dropouts and Log-Softmax layer in the baseline model. We ran both models for 30 epochs. The baseline took 6 hours while CapsNet took 30 hours of real-time computing. One reason behind this was that our pilot study was only allocated one GPU on the cluster and we had to choose a batch size of 5 because of limited memory. We should note that in terms of the number of parameters, the baseline model has  \num[group-separator={,}]{208961829} while CapsNet has \num[group-separator={,}]{124209845} in this training scheme. We reported the computed RMSEs in Table~\ref{RMSE_Tab}. We also show RMSE vs number of epochs in Figure~\ref{fig_rmse} for both training and testing.  As we can see in the results, CapsNet outperformed our baseline model. 

\begin{table}
\centering
 \begin{tabular}{lcc}
  \hline
  Model & Training  & Testing\\
  \hline
  Baseline  & 0.109 & 0.113\\
 CapsNet &  0.081 & 0.103\\
  \hline
 \end{tabular}
 \caption{Computed RMSE between the predictions and true crowd-sourced probabilities; A relative error reduction of 8.8\% was achieved.}
  \label{RMSE_Tab}
\end{table}
 
\subsection{Classification Based on Answers to Question 1 and Reconstruction of Galaxies}\label{Reconst}
In this setup we only chose question one from the question tree, because this question is the most fundamental. Specifically, the question asked ``Is the galaxy simply smooth and rounded, with no sign of a disk?''. There are three answers to this question that determined whether the object is round and smooth (elliptical galaxies), object with disks (spiral galaxies), or an artifact or a star. We first calculated the measure of agreement using equation 7 in \cite{dieleman2015rotation} that reads, 
\begin{equation}
a(p) = 1 - \frac{H(p)}{\log(n)}
\end{equation}
where $H(p) = - \sum_{i=1}^{n}p(x_{i}) \log(p(x_{i}))$ is the entropy of the discrete probability distribution $p(x_{i})$ over $n$ options. The value of $a(p)$ is between 0 and 1 where 0 stands for minimal agreement and 1 stands for maximal agreement. Next, we chose the images where the measure of agreement of participants was $a(p) > 0.8$ where participants only chose between the first two answers (1.1 and 1.2; see Figure 1 and Table 2 in \cite{willett2013galaxy}). For this task, we picked the answer with the highest probability as the correct answer to question one. On 988 images the participants chose 1.1 and on 5,094 images they chose 1.2 as an answer to question 1 with more than 0.8 measure of agreement. 
 
\begin{figure*}
	\centering
 	\subfloat[Training]{\includegraphics[width=0.49\textwidth]{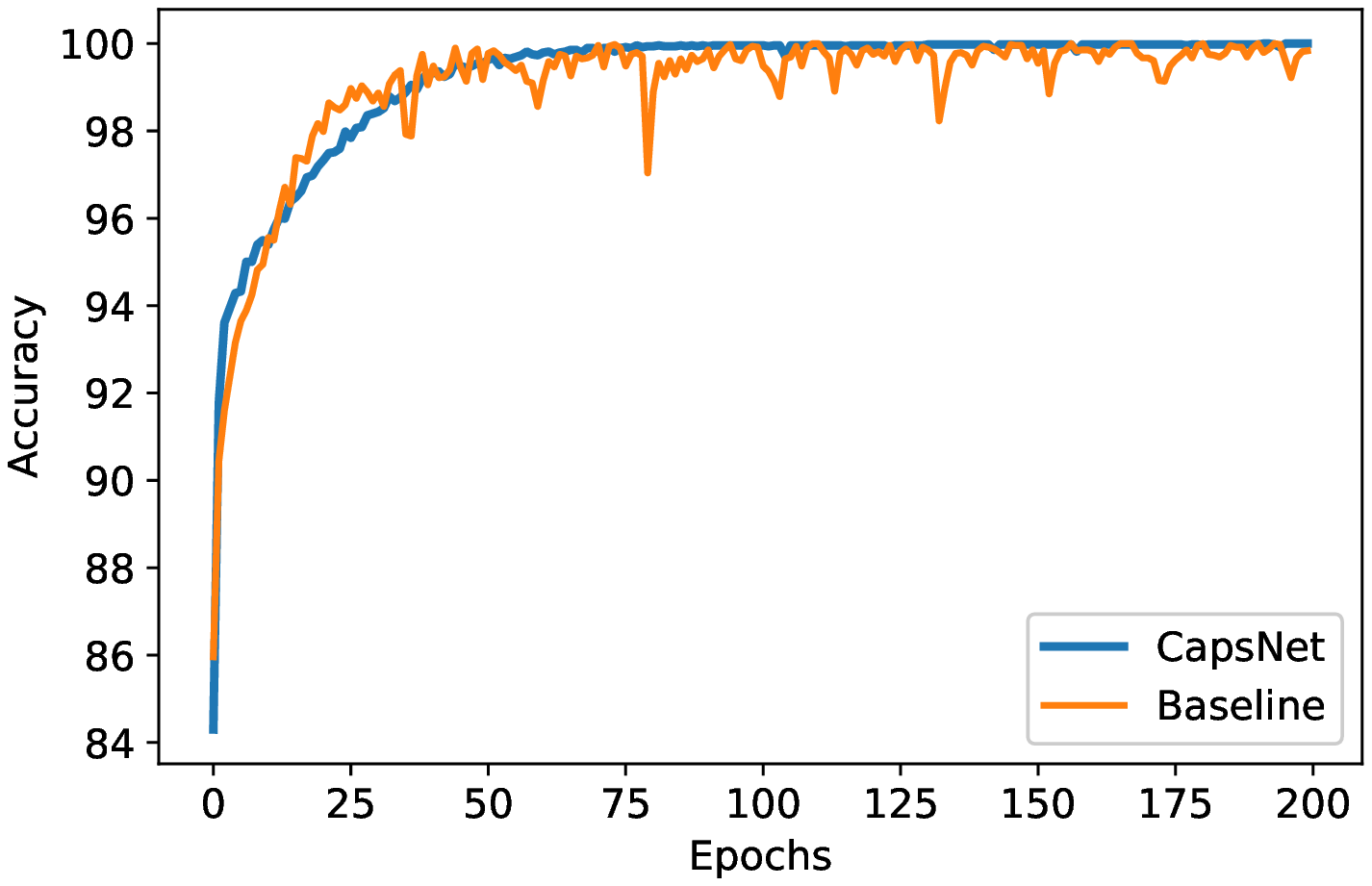}}\hfil 
 	\subfloat[Testing]{\includegraphics[width=0.49\textwidth]{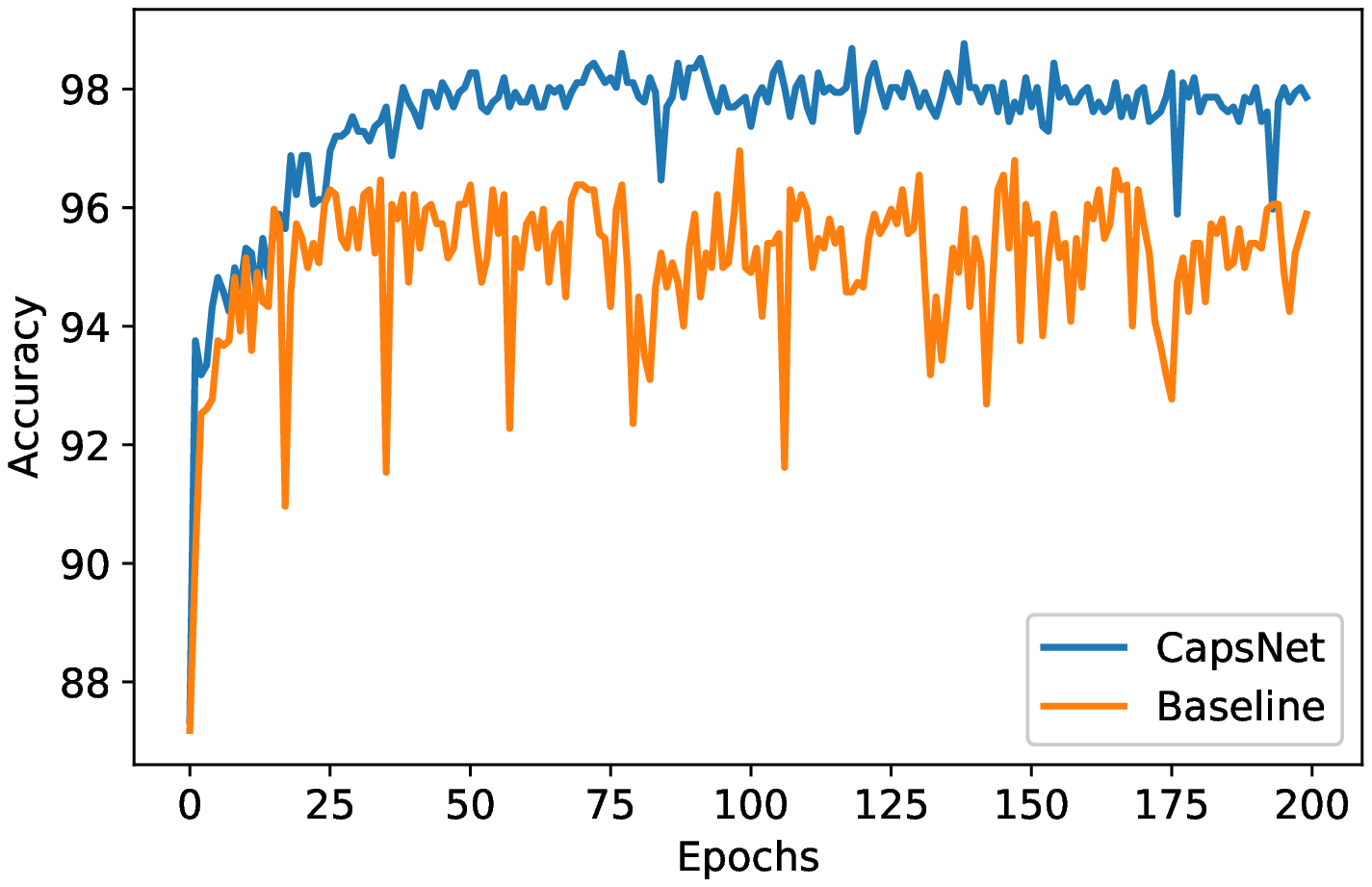}}
    \caption{Training and testing accuracy vs number of epochs for classification based on the answers to question one.}
    \label{fig_3_acc}
\end{figure*}

We trained the baseline model and CapsNet for this scheme for 200 epochs with a batch size of 20 and reported the accuracies in Table~\ref{ACC_Tab}. The training took 1 hour and 3 hours for the baseline model and CapsNet, respectively. In terms of the number of parameters, baseline model had \num[group-separator={,}]{209975554} while CapsNet had \num[group-separator={,}]{28276672} for this training scheme. We show the accuracy curves versus number of epochs for both training and testing in Figure~\ref{fig_3_acc}. As we can see, while the baseline model and CapsNet had similar performance during training, CapsNet outperformed the baseline model at test time. Furthermore, we show the reconstructed images at 10, 100 and 200 epochs generated by CapsNet versus the original images in Figure~\ref{Reconst}. These reconstructed images are very detailed.  

In order to check whether reconstructed images preserved physical properties of the original images, we used brightness profiles of the galaxies to indicate the S{\'e}rsic index \citep{Sersic1963influence} for each galaxy. The S{\'e}rsic profile or the S{\'e}rsic law shows how intensity $I$ of a galaxy changes with the distance $R$ from its center. The S{\'e}rsic profile has the following form,
\begin{equation}
\log(I(R)) = \log\,(I_{0}) - kR^{1/n}
\end{equation}
where $I_{0}$ is the intensity at $R=0$ and $n$ is the S{\'e}rsic index that controls the curvature of the profile. We used the GALFIT software \citep{peng2002detailed} to estimate the S{\'e}rsic index for a subset of reconstructed and original images (116 samples of each) and the results can be found in Figure~\ref{fig_S{\'e}rsic}. We should note that we used a Gaussian Point Spread Function (PSF) with an average FWHM of 6 pixels that was estimated using the stars present in the field. However, \cite{willett2013galaxy} mention that each Galaxy Zoo image was re-scaled to a variable number of arcseconds per pixel during image creation, which causes slight changes in the PSF and therefore GALFIT slightly underestimates or overestimates the S{\'e}rsic index. 

Furthermore, we calculated the mean ($-0.41$) and 95\% confidence interval ($[-1.33, 0.51]$) of the difference between the S{\'e}rsic index estimated for our sample of the original and reconstructed images ($n_{Original} - n_{Reconstructed}$) and the results can be found in Figure~\ref{fig_stat}. These results indicate that the reconstructed images fairly preserved the S{\'e}rsic profile of the original images. However, the estimated S{\'e}rsic index for reconstructed images are mostly larger than the original counterparts. The reason behind this is that the reconstructed images have stronger and spatially larger central light sources than original images; therefore, the estimated S{\'e}rsic indexes are larger for them.

\begin{table}
\centering
 \begin{tabular}{lcc}
  \hline
  Model & Training  & Testing\\
  \hline
  Baseline  & 100\% & 96.96\%\\
 CapsNet &  100\% & 98.77\%\\
  \hline
 \end{tabular}
 \caption{Training and testing accuracy vs number of epochs for classification based on the answers to question one; A relative error reduction of 59.8\% was achieved.}
  \label{ACC_Tab}
\end{table}

\begin{figure*}
	\centering
 	\subfloat[Reconstruction after 10 epochs]{\includegraphics[width=0.49\textwidth]{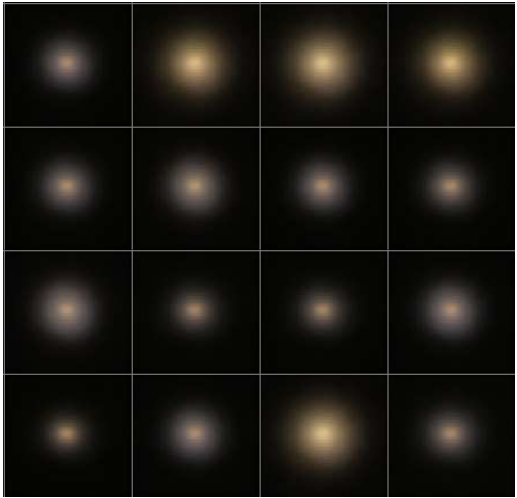}}\hfil 
 	\subfloat[Reconstruction after 100 epochs]{\includegraphics[width=0.49\textwidth]{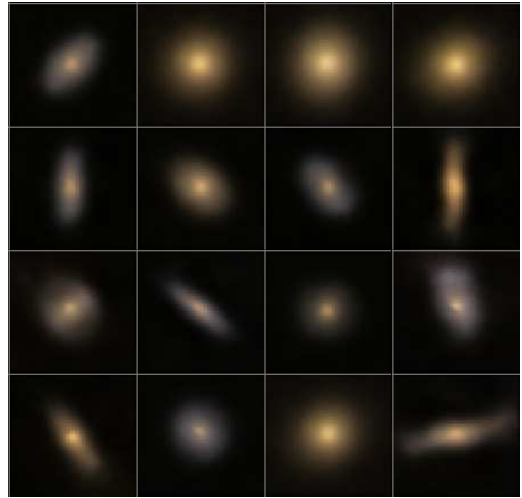}}\\[-2ex]  
 	\subfloat[Reconstruction after 200 epochs]{\includegraphics[width=0.49\textwidth]{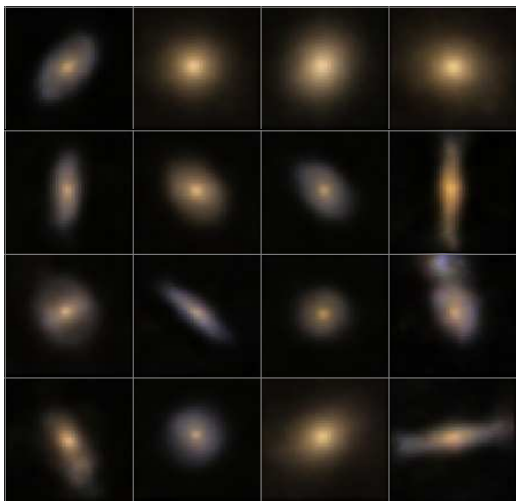}}\hfill
 	\subfloat[Original images]{\includegraphics[width=0.49\textwidth]{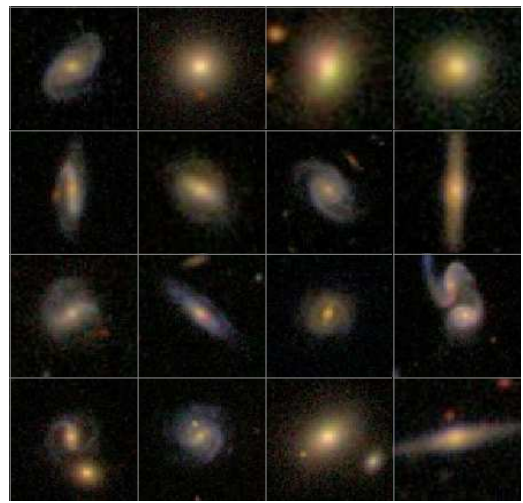}}
 \caption{Original and reconstructed images of the galaxies.}
 \label{Reconst}
\end{figure*}

\begin{figure*}
	\centering
 	\includegraphics[scale = 0.7]{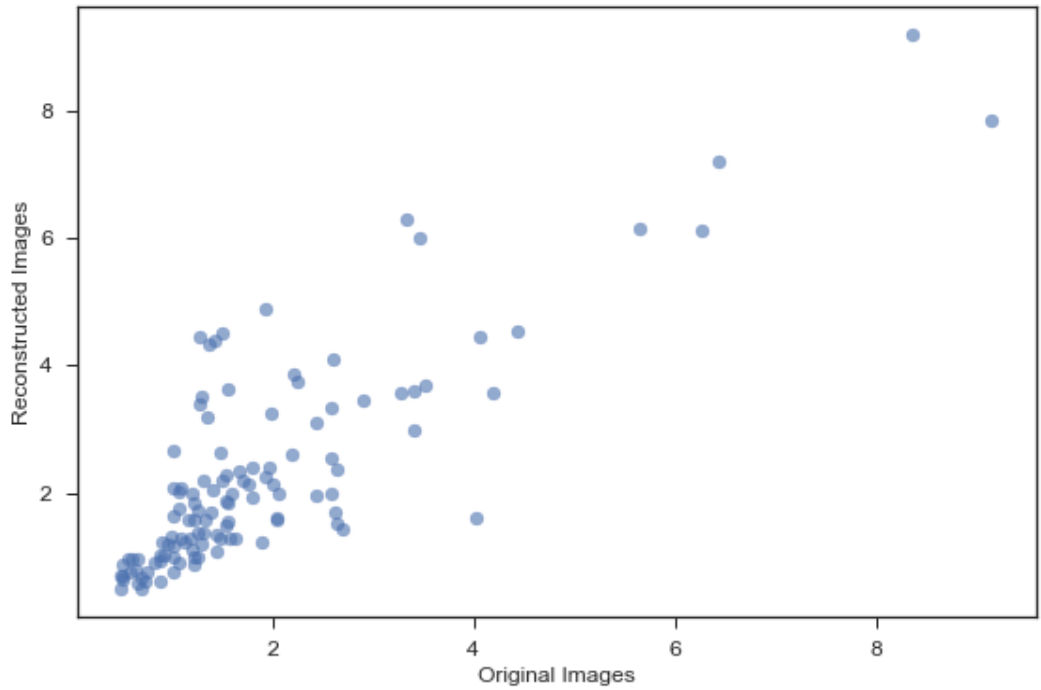} 
    \caption{Estimated S{\'e}rsic index of original images versus reconstructed images.}
    \label{fig_S{\'e}rsic}
\end{figure*}

\begin{figure*}
	\centering
 	\includegraphics[scale = 0.7]{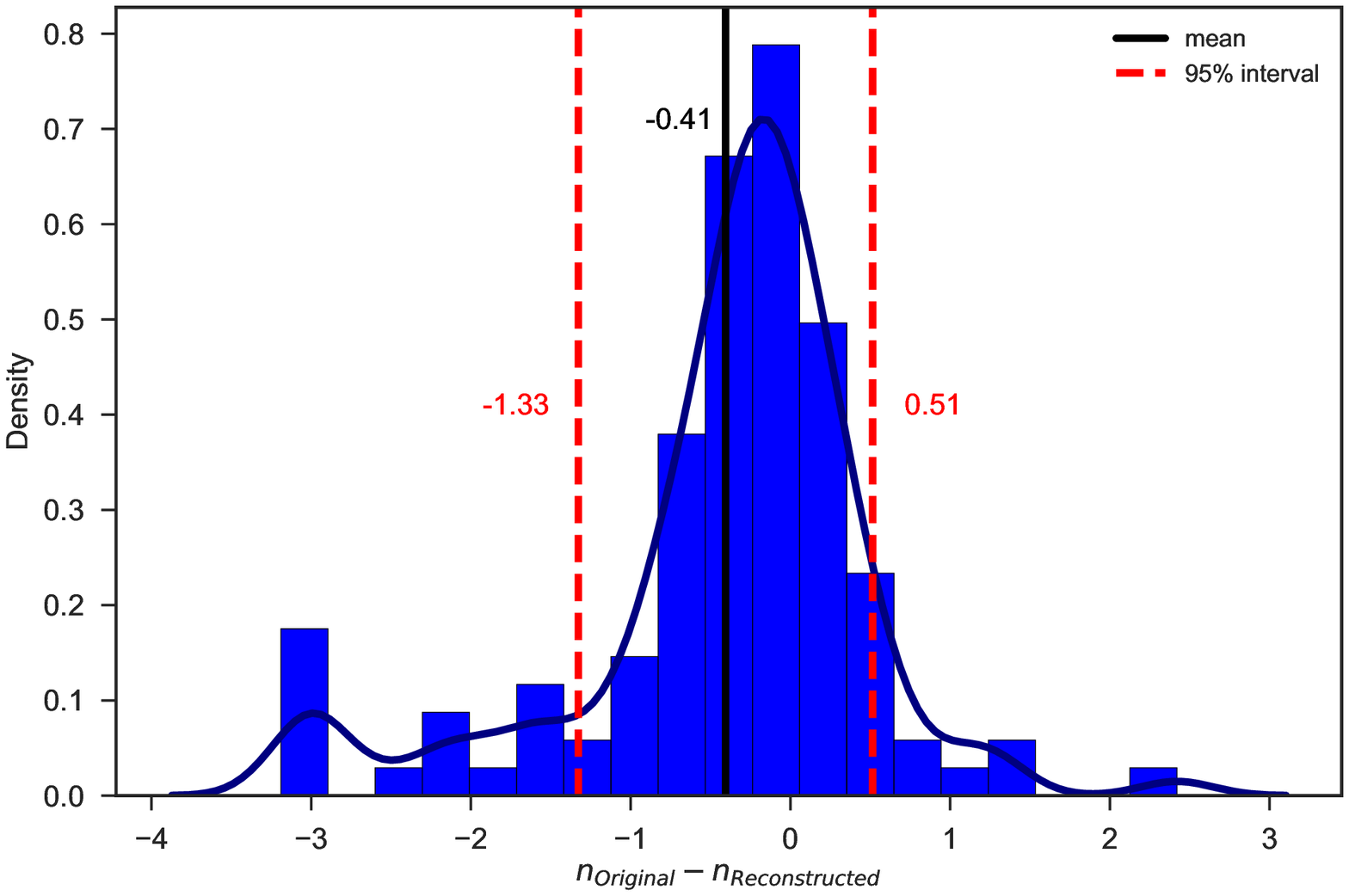} 
    \caption{Mean and 95\% confidence interval of the difference between the estimated S{\'e}rsic index for the original and reconstructed images  ($n_{Original} - n_{Reconstructed} $).}
    \label{fig_stat}
\end{figure*}

\section{Conclusions}

In this work, we presented a new method for performing morphological classification of the galaxies. We used a recently introduced neural network structure called Capsule Network in two different scenarios. 

In the first scenario, we trained models to predict the true crowd-sourced probabilities using both our baseline model and CapsNet. As shown in Table~\ref{RMSE_Tab}, CapsNet clearly outperforms the baseline CNN.

In the second scenario, we chose objects where the participants had more than 0.8 agreement when answering question 1 from the question tree in the Galaxy Zoo project. Next, we chose the answer with the highest probability to be the class of the object. As we can see in Table~\ref{ACC_Tab}, CapsNet outperformed the baseline CNN. We also reconstructed galaxy images using the decoder part of CapsNet that were very detailed and very close to their original counterparts. Furthermore, the S{\'e}rsic index of the galaxies shows that the reconstructed images preserve the physical properties of the original images. However, the estimated S{\'e}rsic index for the reconstructed images is higher than the estimated S{\'e}rsic index of their original counterparts. This can be explained by a larger central light source in the reconstructed images. Thus, training the network on larger datasets with more resolution will be a possible solution to improve this result. 

CapsNet worked really well despite the fact that we did not do any data augmentation and view point extraction similar to \cite{dieleman2015rotation} and our network is much shallower compared to the one presented in their work. Another thing to note is that the CapsNet proposed here has many fewer parameters compared to the baseline CNN. Therefore, we believe that CapsNet is more suitable for the task of galaxy morphological classification. However, we should note that the current implementations of the routing by agreement is slow and more work is needed to reduce the computational complexity. 

We should note that in our work we used the same number of capsules and the same values for $\lambda, m^{+}$ and $m^{-}$ as in \cite{sabour2017dynamic}. In the future, we would like tune the depth of the network and the number of capsules used in the network along with the different values of the parameters. Additionally, it would be interesting to apply CapsNet on larger and more recent datasets generated by the Galaxy Zoo project. Furthermore, extending to multiple GPUs will help to overcome the limitations of our pilot study.  

Upcoming large sky surveys such as LSST will increase the amount of data on galaxies dramatically and an automated method for tasks like morphological classification is highly needed. The method presented here is a possible solution for such tasks.

\section*{Acknowledgements}
This work was supported in part by an allocation of computing time from the Ohio Supercomputer Center. 

Funding for the SDSS and SDSS-II has been provided
by the Alfred P. Sloan Foundation, the Participating Institutions,
the National Science Foundation, the U.S. Department
of Energy, the National Aeronautics and Space Administration,
the Japanese Monbukagakusho, the Max Planck Society,
and the Higher Education Funding Council for England.
The SDSS website is \url{http://www.sdss.org/}.

The SDSS is managed by the Astrophysical Research
Consortium for the Participating Institutions. The Participating
Institutions are the American Museum of Natural
History, Astrophysical Institute Potsdam, University of
Basel, University of Cambridge, Case Western Reserve University,
University of Chicago, Drexel University, Fermilab,
the Institute for Advanced Study, the Japan Participation
Group, Johns Hopkins University, the Joint Institute for
Nuclear Astrophysics, the Kavli Institute for Particle Astrophysics
and Cosmology, the Korean Scientist Group, the
Chinese Academy of Sciences (LAMOST), Los Alamos National
Laboratory, the Max-Planck-Institute for Astronomy
(MPIA), the Max-Planck-Institute for Astrophysics (MPA),
New Mexico State University, Ohio State University, University
of Pittsburgh, University of Portsmouth, Princeton
University, the United States Naval Observatory, and the
University of Washington.

%%%%%%%%%%%%%%%%%%%%%%%%%%%%%%%%%%%%%%%%%%%%%%%%%%

%%%%%%%%%%%%%%%%%%%% REFERENCES %%%%%%%%%%%%%%%%%%

% The best way to enter references is to use BibTeX:

\bibliographystyle{mnras}
\bibliography{bibfile} % if your bibtex file is called example.bib

% Alternatively you could enter them by hand, like this:
% This method is tedious and prone to error if you have lots of references
%\begin{thebibliography}{99}
%\bibitem[\protect\citeauthoryear{Author}{2012}]{Author2012}
%Author A.~N., 2013, Journal of Improbable Astronomy, 1, 1
%\bibitem[\protect\citeauthoryear{Others}{2013}]{Others2013}
%Others S., 2012, Journal of Interesting Stuff, 17, 198
%\end{thebibliography}

%%%%%%%%%%%%%%%%%%%%%%%%%%%%%%%%%%%%%%%%%%%%%%%%%%

%%%%%%%%%%%%%%%%% APPENDICES %%%%%%%%%%%%%%%%%%%%%

%\appendix
%
%\section{Some extra material}
%
%If you want to present additional material which would interrupt the flow of the main paper,
%it can be placed in an Appendix which appears after the list of references.

%%%%%%%%%%%%%%%%%%%%%%%%%%%%%%%%%%%%%%%%%%%%%%%%%%

% Don't change these lines
\bsp	% typesetting comment
\label{lastpage}
\end{document}